\documentclass[10pt]{article}
\begin{document}
\begin{center}
{\large\bf Late-time Inhomogeneity and Acceleration of the
Universe} \vskip 0.3 true in {\large J. W. Moffat} \vskip 0.3 true
in {\it The Perimeter Institute for Theoretical Physics, Waterloo,
Ontario, N2J 2W9, Canada} \vskip 0.3 true in and \vskip 0.3 true
in {\it Department of Physics, University of Waterloo, Waterloo,
Ontario N2L 3G1, Canada}
\end{center}
\begin{abstract}%
The inhomogeneous distribution of matter in the non-linear regime
of galaxies, clusters of galaxies and voids is described by an
exact, spherically symmetric inhomogeneous solution of Einstein's
gravitational field equations, corresponding to an under-dense
void. The solution becomes the homogeneous and isotropic
Einstein-de Sitter solution for a red shift $z > 10-20$, which
describes the matter dominated CMB data with small
inhomogeneities. A spatial volume averaging of physical quantities
is introduced and the averaged time evolution expansion parameter
$\theta$ can give rise in the late-time universe to a volume
averaged deceleration parameter $\langle q\rangle$ that is
negative for a positive matter density. This allows for a region
of accelerated expansion which does not require a cosmological
constant. A negative deceleration parameter can be derived by this
volume averaging procedure from the Lema\^{i}tre-Tolman-Bondi open
void solution, which describes the late-time non-linear regime
associated with galaxies and under-dense voids and solves the
``coincidence'' problem.
\end{abstract}
\vskip 0.2 true in email: john.moffat@perimeterinstitute.ca


\section{Introduction}

In recent articles, we investigated a cosmology in which a
spherically symmetric inhomogeneous enhancement is embedded in an
asymptotic FLRW universe~\cite{Moffat,Moffat2}.\footnote{For a
more complete list of references, see refs.~\cite{Moffat,Moffat2}}
The inhomogeneous enhancement is described by an exact
inhomogeneous solution of Einstein's field equations. The
inhomogeneous distribution of matter at late times can possibly
allow for a negative deceleration parameter $q$, depending on the
nature of the exact inhomogeneous solution of Einstein's field
equations. This will allow for the possibility of explaining the
acceleration of the universe, without a cosmological constant. The
model also leads to an axis pointing towards the center of a
spherically symmetric inhomogeneous enhancement with dipole,
quadrupole and octopole moments aligned with the axis. The
distribution of CMB temperature fluctuations can be unevenly
distributed in the northern and southern
hemispheres~\cite{Moffat}.

We shall describe the inhomogeneous late-time universe by an exact
spherically symmetric solution of Einstein's field equations. The
solution becomes the homogeneous and isotropic Einstein-de Sitter
matter dominated solution for $z> 20$, so that it can describe the
WMAP data near the surface of last scattering with small
temperature perturbations $\delta T/T\sim 10^{-5}$~\cite{Spergel}.
However, for $z < 20$ the solution is inhomogeneous and for
luminosity distances corresponding to $z\sim 0.2-1.4$, includes
the Type SNIa supernova measurements~\cite{Perlmutter,Riess}. We
investigate the expansion geometry of the spherical distribution
of inhomogeneous matter and under-dense voids~\cite{Hoyle} to see
whether it allows for the possibility of a negative $q$ and an
accelerating expansion of the universe, without negative pressure
dark energy or a cosmological constant.

A generic relativistic solution of a three-dimensional
inhomogeneous universe is presently unknown. The highly symmetric
Lema\^{i}tre-Tolman-Bondi (LTB)~\cite{Lemaitre,Tolman,Bondi}
solution that we shall consider in the following is inhomogeneous
in only one spatial dimension. Even though the assumption of
spherical symmetry is unrealistic, the model is useful in
illustrating important {\it local} features of physical
quantities.

For our inhomogeneous late-time universe model, we are required to
carry out a volume averaging of physical scalar quantities, such
as the expansion parameter $\theta$. We find for such an averaging
process that for a zero cosmological constant and for irrotational
matter the average deceleration parameter $\langle q\rangle$ can
be negative and describe an accelerating local region of the
universe, corresponding to our observed Hubble radius, without
invoking a cosmological constant.

\section{Inhomogeneous Friedmann Equations}

For the sake of notational clarity, we write the FLRW line element
\begin{equation}
ds^2=dt^2-a^2(t)\biggl(\frac{dr^2}{1-kr^2}+r^2d\Omega^2\biggr),
\end{equation}
where $k=+1,0,-1$ for a closed, flat and open universe,
respectively, and $d\Omega^2=d\theta^2+\sin\theta^2d\phi^2$. The
general, spherically symmetric inhomogeneous line element is given
by~\cite{Lemaitre,Tolman,Bondi,Bonnor,Moffat3,Moffat4,Krasinski,Moffat}:
\begin{equation}
\label{inhomometric} ds^2=dt^2-X^2(r,t)dr^2-R^2(r,t)d\Omega^2.
\end{equation}
The energy-momentum tensor ${T^\mu}_\nu$ takes the barytropic form
\begin{equation}
\label{energymomentum} {T^\mu}_\nu=(\rho+p)u^\mu u_\nu
-p{\delta^\mu}_\nu,
\end{equation}
where $u^\mu=dx^\mu/ds$ and, in general, the density
$\rho=\rho(r,t)$ and the pressure $p=p(r,t)$ depend on both $r$
and $t$. We have for comoving coordinates $u^0=1, u^i=0,\,
(i=1,2,3)$ and $g^{\mu\nu}u_\mu u_\nu=1$.

The Einstein gravitational equations are
\begin{equation}
\label{Einstein} G_{\mu\nu}+\Lambda g_{\mu\nu}=-8\pi GT_{\mu\nu},
\end{equation}
where $G_{\mu\nu}=R_{\mu\nu}-\frac{1}{2}g_{\mu\nu}{\cal R}$,
${\cal R}=g^{\mu\nu}R_{\mu\nu}$ and $\Lambda$ is the cosmological
constant. Solving the $G_{01}=0$ equation for the metric
(\ref{inhomometric}), we find that
\begin{equation}
X(r,t)=\frac{R'(r,t)}{f(r)},
\end{equation}
where $R'=\partial R/\partial r$ and $f(r)$ is an arbitrary
function of $r$.

We obtain the two generalized Friedmann equations~\cite{Moffat}:
\begin{equation}
\label{inhomoFriedmann} \frac{{\dot R}^2}{R^2}+2\frac{{\dot
R}'}{R'}\frac{{\dot R}}{R}+\frac{1}{R^2}(1-f^2)
-2\frac{ff'}{R'R}=8\pi G\rho+\Lambda,
\end{equation}
\begin{equation}
\label{inhomoFriedmann2} \frac{\ddot
R}{R}+\frac{1}{3}\frac{\dot{R}^2}{R^2}
+\frac{1}{3}\frac{1}{R^2}(1-f^2) -\frac{1}{3}\frac{{\dot
R}'}{R'}\frac{{\dot R}}{R}+\frac{1}{3} \frac{ff'}{R'R}
=-\frac{4\pi G}{3}(\rho+3p)+\frac{1}{3}\Lambda,
\end{equation}
where $\dot R=\partial R/\partial t$.

\section{Late-Time Matter Dominated Universe}

The late-time matter dominated universe will be pictured as a
large-scale inhomogeneous enhancement that is described by an
exact inhomogeneous spherically symmetric solution of Einstein's
field equations. The inhomogeneous enhancement is embedded in a
matter dominated universe that approaches asymptotically an
Einstein-de Sitter universe as $t\rightarrow\infty$. An observer
will be off-center from the origin of coordinates of the spherical
inhomogeneous enhancement. The local inhomogeneous solution
corresponds to an open, hyperbolic solution and describes an
under-dense void. Surveys such as the 2-degree Field Galaxy
Redshift Survey and the Sloan Digital Sky Survey show large
volumes of relatively empty space, or voids, in the distribution
of galaxies~\cite{Hoyle}.

For the matter dominated Lema\^{i}tre-Tolman-Bondi
(LTB)~\cite{Lemaitre,Tolman,Bondi} model with zero pressure $p=0$
and zero cosmological constant $\Lambda=0$, the Einstein field
equations demand that $R(r,t)$ satisfies
\begin{equation}
\label{Requation} 2R(r,t){\dot R}^2(r,t)+2R(r,t)(1-f^2(r))=F(r),
\end{equation}
with $F$ being an arbitrary function of $r$ of class $C^2$.

The proper density of matter can be expressed as
\begin{equation}
\label{density} \rho=\frac{F'}{16\pi GR'R^2}.
\end{equation}
We can solve (\ref{density}) to obtain
\begin{equation}
\Omega-1\equiv\frac{\rho}{\rho_c}-1=\frac{1}{H^2_{\rm
eff}}\biggl(\frac{1-f^2}{R^2}-2\frac{f}{R}\frac{f'}{R'}\biggr),
\end{equation}
where
\begin{equation}
H^2_{\rm eff}=H^2_\perp+2H_\perp H_r,
\end{equation}
is an effective Hubble parameter and
\begin{equation}
H_r=\frac{{\dot R}'}{R'},\quad H_\perp=\frac{{\dot R}}{R}.
\end{equation}
We have for the critical density
\begin{equation}
8\pi G\rho_c=H^2_{\rm eff}.
\end{equation}
There are three possibilities for the curvature of spacetime: 1)
$f^2 > 1$ open ($\Omega-1 < 0$), 2) $f^2=1$ flat ($\Omega-1=0$),
$f^2 < 1$ closed ($\Omega-1 > 0$).

We describe the inhomogeneous density regime by a hyperbolic, open
solution $f^2(r) > 1$ with $p=\Lambda=0$, corresponding to an
under-dense void. We choose
\begin{equation}
f(r)=\sqrt{1+r^2}
\end{equation}
and the metric reduces to
\begin{equation}
\label{openmetric}
ds^2=dt^2-\frac{R^{'2}(r,t)}{1+r^2}dr^2-R^2(r,t)d\Omega^2.
\end{equation}
A parametric solution is given by
\begin{equation}
\label{inhomoopen} R(r,t)=\frac{1}{4}F(r)(f^2(r)-1)^{-1}(\cosh
u(r,t)-1),
\end{equation}
\begin{equation}
t+\beta(r)=\frac{1}{4}(f^2(r)-1)^{-3/2}(\sinh u(r,t)-u(r,t)),
\end{equation}
\begin{equation}
8\pi G\rho(r,t)
=\biggl(\frac{2F'(r)}{F^2(r)}\biggr)\biggl(\frac{f^2(r)-1}{R'(r,t)}\biggr)
\biggl(\frac{1}{\sinh^4\frac{1}{2}u(r,t)}\biggr).
\end{equation}
Here, $\beta(r)$ is an arbitrary function of $r$ of class
$C^2$~\cite{Bonnor}.

The function $\beta(r)$ can be specified in terms of a density on
some spacelike hypersurface $t=t_0$. The metric and density are
singular for
\begin{equation}
u=0\quad {\rm or}\quad R'(r,t)=0,
\end{equation}
where $u=0$ refers to the hypersurface $t+\beta(r)=0$ and the
singular surface for $R'(r,t)=0$ is more complicated. Our
pressureless model requires that the singular surface
$t(r)=\Sigma(r)$ describes the surface on which the universe
becomes matter dominated (in the FLRW model this occurs at $z\sim
10^4$). The hyperbolic FLRW model is obtained by choosing the
conditions
\begin{equation}
f^2(r)=1+r^2,\quad F(r)=4sr^3,\quad \beta(r)=0,
\end{equation}
where $s$ is a positive constant. This leads to the metric
solution
\begin{equation}
R=sr(\cosh u-1),
\end{equation}
\begin{equation}
t=s(\sinh u-u),
\end{equation}
\begin{equation}
8\pi
G\rho=\frac{3}{4s^2}\biggl(\frac{1}{\sinh^6\frac{1}{2}u}\biggr).
\end{equation}

We shall change the radial coordinate to
\begin{equation}
r^*=r-r_{SLS},
\end{equation}
where $r=r_{SLS}$ is the location of the surface of last
scattering corresponding to $r^*=0$ at a red shift $z\sim 1100$.
We postulate that $\beta(r^*)$ approaches a maximum constant value
at the approximate end of galaxy and cluster formation, $z_f\sim
1-5$, where $z=z_f$ denotes the final value of the red shift. We
assume that $\beta(r^*)\rightarrow 0$ and $f(r^*)\rightarrow 1$ as
$r^*\rightarrow 0$. A model for $\beta$ is of the form:
\begin{equation}
\beta(r^*)=a\biggl(\frac{r^*}{b+r^*}\biggr)^\alpha,
\end{equation}
where $a$, $b$ and $\alpha$ are positive constants, $\beta(0)=0$
and $\beta(r^*)\rightarrow a$ for $r^* \gg b$. The surface $r^*=0$
for the metric is a singular surface.

The metric for $f(r^*)\rightarrow 1$ becomes~\cite{Bonnor}:
\begin{equation}
\label{matterdommetric}
ds^2=dt^2-(t+\beta(r^*))^{4/3}(Y^2(r^*,t)dr^{*2}+r^{*2}d\Omega^2),
\end{equation}
where
\begin{equation}
Y(r^*,t)=1+\frac{2r^*\beta'(r^*)}{3(t+\beta(r^*))},
\end{equation}
and
\begin{equation}
\rho(r^*,t)=\frac{1}{6\pi G(t+\beta(r^*))^2Y(r^*,t)}.
\end{equation}
We obtain for $\beta(r^*)\rightarrow 0$ the homogeneous and
isotropic solution corresponding to a spatially flat Einstein-de
Sitter universe with the metric
\begin{equation}
ds^2=dt^2-a^2(t)(dr^{*2}+r^{*2}d\Omega^2),
\end{equation}
and $a(t)\propto t^{2/3}$. This solution is compatible as a
background spacetime with the approximately homogeneous and
isotropic WMAP data for $20 < z < 10^3$ and $\delta\rho/\rho\sim
10^{-5}$~\cite{Spergel}.

The Einstein-de Sitter solution must be supplemented by non-zero
pressure contributions for $z\sim 10^3$ in the linear regime at
the surface of last scattering in order to take into account the
radiation density $\rho_R$. Moreover, a more general solution is
required that has a non-vanishing cosmological constant $\Lambda$,
so that we have $\Omega=\Omega_M+\Omega_R +\Omega_\Lambda=1$ in
agreement with the spatially flat result $\Omega=1.02\pm 0.02$,
obtained from the determination of the first acoustic peak in the
WMAP power spectrum~\cite{Spergel}. However, in the following
section, we shall replace $\Omega_\Lambda$ by a contribution
corresponding to an acceleration parameter $q$ obtained from our
inhomogeneous void solution with $\Lambda=0$.

\section{The Inhomogeneous Late-Time Density and Deceleration Parameters}

We obtain from (\ref{inhomoFriedmann}) for $f(r)=\sqrt{1+r^2}$ and
$p=0$ (we replace for convenience $r^*$ by $r$):
\begin{equation}
H_\perp^2+2H_rH_\perp-\frac{r^2}{R^2}-\frac{2r}{R'R}=8\pi
G\rho+\Lambda.
\end{equation}
Dividing this equation by $H_\perp^2$ gives
\begin{equation}
\label{omegadensity} \Omega\equiv\frac{8\pi
G\rho}{H_\perp^2}+\frac{\Lambda}{H^2_\perp}+\frac{r^2}{R^2H_\perp^2}
+\frac{2r}{R'RH_\perp^2}-\frac{2H_r}{H_\perp}=1.
\end{equation}

Let us expand $R(r,t)$ in a Taylor series
\begin{equation}
R(r,t)=R[r,t_0-(t_0-t)] =R(r,t_0)\biggl[1-(t_0-t)\frac{{\dot
R}(r,t_0)}{R(r,t_0)} +\frac{1}{2}(t_0-t)^2\frac{{\ddot
R}(r,t_0)}{R(r,t_0)} -...\biggr]
$$ $$
=R(r,t_0)\biggl[1-(t_0-t)H_{0\perp}-\frac{1}{2}(t_0-t)^2q(r,t_0)H_{0\perp}^2-...\biggr],
\end{equation}
where $t_0$ denotes the present epoch and $H_{0\perp}={\dot
R}(r,t_0)/R(r,t_0)$. Moreover, we have
\begin{equation}
q(r,t_0)=-\frac{{\ddot R}(r,t_0)R(r,t_0)}{{\dot R}^2(r,t_0)}.
\end{equation}
For the matter dominated era we set $p=0$ in
Eqs.(\ref{inhomoFriedmann}) and (\ref{inhomoFriedmann2}) and
substitute for ${\ddot R}$ from (\ref{inhomoFriedmann2}) to obtain
\begin{equation}
\label{inhomodeceleration}
q(r,t_0)=\frac{1}{3}+\frac{4\pi\rho_0(r,t_0)}{3H_{0\perp}^2(r,t_0)}
$$ $$
+\frac{1}{3}\frac{r}{RR'H^2_{0\perp}}
-\frac{\Lambda}{3H^2_{0\perp}(r,t_0)}-\frac{1}{3}\frac{H_{0r}(r,t_0)}{H_{0\perp}(r,t_0)}
-\frac{1}{3}\frac{r^2}{R^2H^2_{0\perp}},
\end{equation}
where $H_{0r}(r,t_0)={\dot R}'(r,t_0)/R'(r,t_0)$.

If we set $H_{0\perp}(r,t_0)=H_{0r}(r,t_0)=H(t_0)$ where
$H(t_0)={\dot a}(t_0)/a(t_0)$ and $R(r,t_0)=a(t_0)$, then we
obtain the {\it global} FLRW expression for the deceleration
parameter for a spatially flat universe with $f^2(r)=1$:
\begin{equation}
q_0=\frac{1}{2}\Omega_{0M}-\Omega_{0\Lambda},
\end{equation}
where from (\ref{omegadensity}) $\Omega_{0M}+\Omega_{0\Lambda}=1$
with $\Omega_{0M}=8\pi\rho_{0M}/3H^2(t_0)$ and
$\Omega_{0\Lambda}=\Lambda/3H^2(t_0)$.

The variance of $q$ is given by the exact non-perturbative
expression:
\begin{equation}
{\rm var}(q)\equiv\langle q^2-\langle
q\rangle^2\rangle^{1/2}=\overline{q},
\end{equation}
where
\begin{equation}
\label{spatialaverage} \overline{(...)}=\frac{\int
d^3x\sqrt{\gamma}(...)}{\int d^3x\sqrt{\gamma}},
\end{equation}
denotes the ensemble average and $\gamma$ denotes the determinant
of the 3-dimensional metric $g_{ij}\,(i=1,2,3)$.

We see from (\ref{inhomodeceleration}) that different observers
located in different causally disconnected parts of the sky will
observe different values for the deceleration parameter $q$,
depending upon their location and distance from the center of the
spherically symmetric inhomogeneous enhancement. This can lead to
one form of cosmic variance, because the spatial average of all
the observed values of local physical quantities, including the
deceleration parameter, $q$, will have an intrinsic uncertainty.

Let us set the cosmological constant to zero, $\Lambda=0$
\footnote{We do not provide any solution to the cosmological
constant problem, namely, why $\Lambda=0$.}. From
(\ref{omegadensity}) and (\ref{inhomodeceleration}), we obtain for
$t=t_0$:
\begin{equation}
\label{density2} \Omega_0\equiv \frac{8\pi
G\rho_0}{H^2_{0\perp}}+\frac{r^2}{R^2H^2_{0\perp}}
+\frac{2r}{R'RH^2_{0\perp}}-\frac{2H_{0r}}{H_{0\perp}}=1,
\end{equation}
and
\begin{equation}
\label{inhomodeceleration2} q(r,t_0)=\frac{1}{3} +\frac{4\pi
G\rho_0(r,t_0)}{3H^2_{0\perp}(r,t_0)}
+\frac{1}{3}\frac{r}{RR'H^2_{0\perp}}
-\frac{1}{3}\frac{H_{0r}(r,t_0)}{H_{0\perp}(r,t_0)}-\frac{1}{3}\frac{r^2}{R^2H^2_{0\perp}}.
\end{equation}

If in (\ref{inhomodeceleration2}) we have
\begin{equation}
\label{accelcond} H_{0r}H_{0\perp}+\frac{r^2}{R^2} > 4\pi
G\rho_0+H^2_{0\perp}+\frac{r}{RR'},
\end{equation}
then the deceleration parameter $q$ can be negative {\it and cause
the universe to accelerate without a cosmological constant or dark
energy}. If we solve for $H_{0r}/H_{0\perp}$ from
Eq.(\ref{density2}) and substitute into
(\ref{inhomodeceleration2}), we obtain
\begin{equation}
\label{qacceleration}
q(r,t_0)=\frac{1}{2}-\frac{5}{6}\biggl(\frac{r^2}{R^2H^2_{0\perp}}\biggr),
\end{equation}
which gives a negative $q$ for
\begin{equation}
\label{accelcond2} r^2/R^2H^2_{0\perp}> 3/5,
\end{equation}
and we can satisfy the condition (\ref{density2}). We see that for
$r=0$ corresponding to $f^2(r)=1$ and a spatially flat universe
$q(r,t_0)=1/2$. For $p=\Lambda=0$, Eqs.(\ref{inhomoFriedmann2})
and (\ref{accelcond}) lead to ${\ddot R}> 0$ as the condition for
an accelerating expansion of the universe.

The conditions (\ref{density2}) and (\ref{accelcond}) require that
$f^2(r)> 1$ for the inhomogeneous enhancement, corresponding to an
under-dense void. If we choose instead to describe the
inhomogeneity by the spatially flat solution $f^2(r)=1$, then we
cannot simultaneously satisfy $\Omega=1$ and $q <0$. However, we
will investigate in the next Section, whether the condition
(\ref{accelcond}) can indeed be satisfied for the particular LTB
model that we have adopted. We shall find that a more general
inhomogeneous solution is needed to establish whether such a
solution can explain the accelerating expansion of the universe
without a negative pressure fluid or a cosmological constant.

\section{The Raychaudhuri Equation}

The Raychaudhuri equation is~\cite{Raychaudhuri}
\begin{equation}
\label{Ray} \frac{d\theta}{ds}=-R_{\mu\nu}V^\mu
V^\nu+2\omega^2-2\sigma^2-\frac{1}{3}\theta^2,
\end{equation}
where we have adopted geodesic world lines, $V^\mu$ is a time-like
vector and
\begin{equation}
\omega^2=\omega^{\mu\nu}\omega_{\mu\nu} \geq 0,\quad
\sigma^2=\sigma^{\mu\nu}\sigma_{\mu\nu}\geq 0.
\end{equation}
Here, $\sigma_{\mu\nu}$ and $\omega_{\mu\nu}$ denote the shear
tensor and the vorticity tensor, respectively. From Einstein's
field equations, we have for $p=\Lambda=0$:
\begin{equation}
\label{Ray2}
\frac{d\theta}{ds}=2\omega^2-2\sigma^2-\frac{1}{3}\theta^2-4\pi
G\rho.
\end{equation}
The deceleration parameter $q$ is defined by
\begin{equation}
q\equiv
-\frac{(3d\theta/ds+\theta^2)}{\theta^2}=\frac{6(\sigma^2-\omega^2)+4\pi
G\rho}{\theta^2}.
\end{equation}
If the vorticity $\omega=0$, then the deceleration parameter $q$
has to be positive for positive $\rho$.

For our synchronous comoving coordinates we have
\begin{equation}
\frac{d\theta}{ds}=V^\mu\nabla_\mu\theta=V^\mu\partial_\mu\theta={\dot\theta}.
\end{equation}
We have
\begin{equation}
q\equiv
-\frac{(3\dot\theta+\theta^2)}{\theta^2}=-\frac{\biggl(3\frac{{\ddot
R}'}{R'}+6\frac{{\ddot
R}}{R}+\theta^2-3H_r^2-6H_\perp^2\biggr)}{\theta^2}.
\end{equation}
We may now conclude that if
\begin{equation}
\label{qaccel} 3H_r^2+6H_\perp^2 > 3\frac{{\ddot
R}'}{R'}+6\frac{{\ddot R}}{R}+\theta^2,
\end{equation}
then $q$ could be negative leading to a local acceleration of the
universe. However, for our synchronous comoving frame, we have
\begin{equation}
\omega_{\mu\nu}\equiv \nabla_{[\nu}V_{\mu]}=\partial_\nu
V_\mu-\partial_\mu V_\nu=0,
\end{equation}
so that the vorticity for this chosen gauge is zero. Therefore, we
must deduce that the condition (\ref{qaccel}) cannot be satisfied
for the positive energy condition $\rho > 0$ and $p > 0$.
Moreover, the two conditions (\ref{accelcond}) and
(\ref{accelcond2}) cannot be satisfied for the synchronous and
comoving metric (\ref{inhomometric}). However, in the following
section, we will show that we must carry out a spatial volume
averaging of the expansion parameter $\theta$ and its time
evolution to arrive at a physically viable description of the
local inhomogeneous late-time universe and an accelerating
expansion of the universe.

\section{Spatially Averaged Cosmological Domains}

For our inhomogeneous model, it is necessary to perform a spatial
average (\ref{spatialaverage}) of physical quantities, due to
their observer, location dependence. Let us define a  more
specific domain averaged expression for a scalar
quantity~\cite{Ellis,Buchert,Kolb}:
\begin{equation}
\langle\Psi({\vec x},t)\rangle_D=\frac{1}{{\cal V_D}}\int_D
d^3x\sqrt{\gamma}\Psi({\vec x},t),
\end{equation}
where
\begin{equation}
{\cal V}_D=\int_Dd^3x\sqrt{\gamma}
\end{equation}
is the volume of the simply-connected domain, $D$, in a
hypersurface. We can define an effective scale-factor for our
spatially averaged spherically symmetric inhomogeneous enhancement
(void):
\begin{equation}
\langle R(r,t)\rangle_D=\biggl(\frac{{\cal V}(t)_D}{{\cal
V}_{iD}}\biggr)^{1/3},
\end{equation}
where ${\cal V}_{iD}$ is the initial volume.

The volume averaging of the scalar $\Psi$ does not commute with
its time evolution~\cite{Ellis,Buchert,Kolb}:
\begin{equation}
\langle\dot{\Psi}(r,t)\rangle_D-\partial_t\langle\Psi(r,t)\rangle_D
=\langle\Psi(r,t)\rangle_D\langle\theta(r,t)\rangle_D
-\langle\Psi(r,t)\theta(r,t)\rangle_D.
\end{equation}
We can derive for the averaged $\langle\theta\rangle_D$ the
equation
\begin{equation}
\langle\theta\rangle_D=3\frac{\langle{\dot R}\rangle_D}{\langle
R\rangle_D}=3\langle H_{\perp}\rangle_D.
\end{equation}
Setting $\Psi=\theta$ and substituting for the Raychaudhuri
equation for an irrotational matter dominated late-time model with
$\omega_{\mu\nu}=0$, $d\theta/ds=d\theta/dt$ and $\Lambda=0$, we
obtain the expression
\begin{equation}
3\frac{\langle{\ddot R}\rangle_D}{\langle R\rangle_D}=-4\pi
G\langle\rho\rangle_D+\langle P\rangle_D,
\end{equation}
where $\langle P\rangle_D$ is a function of
$\langle\theta\rangle_D$, the average Hubble expansion parameters
$\langle H_\perp\rangle_D$ and $\langle H_r\rangle_D$ and the
average shear $\langle\sigma\rangle_D$.

For inhomogeneous cosmology, the smoothing due to averaging of the
Einstein field equations does not commute with the time evolution
of the non-linear field equations. This leads to extra
contributions in the effective, averaged Einstein field equations,
which do not satisfy the usual energy conditions even though they
are satisfied by the original energy-momentum tensor. It is the
lack of commutativity of the time evolution of the expansion of
the universe in a local patch inside our Hubble horizon that
circumvents the no-go theorem based on the local Raychaudhuri
equation~\cite{Moffat2}.

If we can satisfy the condition
\begin{equation}
\label{Pcondition} \langle P\rangle_D >  4\pi
G\langle\rho\rangle_D,
\end{equation}
then the averaged deceleration parameter $\langle q\rangle_D$ can
be negative corresponding to the acceleration of a local patch of
the universe. If we substitute the value of $\langle{\ddot
R}\rangle_D$ from the volume averaged equation
(\ref{inhomoFriedmann2}) for $\Lambda=0$, then the exact LTB
solution for an {\it irrotational} inhomogeneous, under-dense void
solution can describe the local acceleration of the late-time
universe and fulfill the constraint equation (\ref{Pcondition}).
The spatial domain average of $q$ in Eq.(\ref{qacceleration})
gives
\begin{equation}
\label{qaccelerationaverage} \langle
q\rangle_D=\frac{1}{2}-\frac{5}{6}\left<\biggl(\frac{r^2}{R^2H^2_{0\perp}}\biggr)\right>_D.
\end{equation}
This yields the averaged condition for an accelerating universe:
\begin{equation}
\left<\frac{r^2}{R^2H^2_{0\perp}}\right>_D > 3/5.
\end{equation}
This spatial domain averaged condition for our LTB void solution
can be satisfied in our synchronous comoving gauge with zero
vorticity for $\langle\rho\rangle_D > 0$.

\section{Conclusions}

We have modelled the late-time inhomogeneous and non-linear regime
for $z < 10$ by an exact spherically symmetric, inhomogeneous
solution of Einstein's field equations. The solution approaches an
Einstein-de Sitter matter dominated solution for $z > 10$, which
describes a homogeneous and isotropic FLRW background spacetime
with small inhomogeneities $\delta\rho/\rho\sim 10^{-5}$ near the
surface of last scattering in agreement with the WMAP
data~\cite{Spergel}. A volume averaging of physical quantities is
necessitated for our inhomogeneous universe, and due to the lack
of commutativity of the time evolution of the average of the
scalar expansion $\theta$, we find that it is possible to have a
negative averaged deceleration parameter $\langle q\rangle$ for a
positive density of matter $\rho$ and a zero vorticity $\omega=0$.
Observational bounds on the magnitude of the quantity $\langle
P_D\rangle$ in the condition (\ref{Pcondition}) must be obtained
to establish whether a sufficiently negative $q$ can be produced
by the expanding inhomogeneous enhancement or void to allow an
explanation of the accelerating universe without a negative
pressure fluid or a cosmological constant.

The inhomogeneous solution is exact, so that we do not have to be
concerned with the failure of perturbative backreaction
calculations in the non-linear late-time universe regime. However,
we have assumed a high degree of symmetry for our inhomogeneous
solution, since it only allows for a one-dimensional inhomogeneity
in the radial direction, although the angular azimuthal expansion
described by $H_\perp$ does play a significant role in determining
the cosmological solution. It would be interesting to investigate
inhomogeneous cosmological solutions that possess less symmetry
than the LTB models~\cite{Krasinski}.

\vskip 0.2 true in {\bf Acknowledgments} \vskip 0.2 true in This
work was supported by the Natural Sciences and Engineering
Research Council of Canada. I thank Joel Brownstein and Martin
Green for helpful discussions.\vskip 0.5 true in

\end{document}